\documentclass[twocolumn,showpacs,preprintnumbers,nofootinbib,amsmath,amssymb]{revtex4}
\usepackage{amsmath,amssymb,amsfonts}
\usepackage{graphicx}
\usepackage{epsfig}

\begin{document}
\title{Pairing in Nuclear Matter and Finite Nuclei}
\author{H. M{\"u}ther}
\email{herbert.muether@uni-tuebingen.de}
 \affiliation{Institute for Theoretical Physics, University of T\"ubingen, Auf der
Morgenstelle 14, D-72076 T\"ubingen, Germany}
\author{A. Polls}
\email{artur@fqa.ub.edu}
\affiliation{Departament  de  F\'\i sica  Qu\`antica  i  Astrof\'\i sica  and  Institut  de  Ci\`encies  del  Cosmos  (ICCUB),
Universitat  de  Barcelona,  E-08028  Barcelona,  Spain}
\begin{abstract}
Effects of pairing with isospin $T=0$ and $T=1$ are systematically studied in a model, which is based on a realistic nucleon-nucleon interaction and allows to describe the transition from infinite nuclear matter to finite nuclei. Special attention is paid to the development of the spin-orbit term in the mean field of nucleons in finite nuclei. The spin-orbit term yields a drastic suppression of $T=0$ proton-neutron pairing but does not lead to a complete disappearance in finite nuclei. Arguments are presented, why no clear evidence of $T=0$ pairing can be observed in the binding energies of finite nuclei.
\end{abstract}
 \maketitle

 \section{Introduction} For several decades a lot of experimental evidence has been accumulated for pairing between nucleons of the same isospin in nuclei. The pairing term is an important ingredient of the nuclear mass formula to describe the odd-even mass staggering in the binding energies of nuclei\cite{ring}. Also other nuclear properties as deformation and moments of inertia deduced from rotational bands can be described by allowing for the isospin vector ($T=1$) pairing in nuclear structure studies based on Hartree-Fock plus BCS or Hartree-Fock Bogoliubov calculations\cite{brink2005,broglia2012,dean2003,fraundorfer14}. 

While the importance of proton-proton ($pp$) and neutron-neutron ($nn$) pairing correlations is established, no clear evidence has empirically been observed for corresponding $pn$ correlations\cite{fraundorfer14,macchia1,isk2018}. At first sight this is rather astonishing since the proton-neutron interaction is more attractive than the interaction between like nucleons and leads to the only vacuum bound states of two nucleons in the deuteron channel $^3SD_1$. Indeed, 
BCS calculations for infinite nuclear matter\cite{baldo90,elgor90,kuckei} predict values for the gap in the
$^1S_0$ channel of the order of 1 to 2 MeV, which is in reasonable agreement with empirical data for $nn$ and $pp$ pairing in finite nuclei. 
Corresponding calculations for $pn$ pairing in the $^3SD_1$ channel yield much larger values for the pairing gap, 
which are of the order of 10 MeV \cite{alm90,vonder91,baldo92,takats93,baldo95,dressedpair05,dressedp06,dressedp07,rubts17}. Therefore one may expect that pairing  should also
be seen in finite nuclei, in particular in light nuclei with equal number of protons and neutrons.

Efforts have been made to determine the $pn$ pairing in nuclei by solving the corresponding Hartree-Fock 
Bogoliubov equations\cite{wolter71} or to extract corresponding correlations from wave-function of shell-model calculations
in one or two major shells\cite{langanke96,vampir}. Neither these theoretical studies nor the analysis of empirical data,
as discussed above, provide any clear evidence for strong $pn$ pairing effects like those occurring in infinite nuclear matter.

It has been argued\cite{poves98,bertsch10,bertsch11} that the strong spin-orbit field in light nuclei may spoil the iso-scalar spin 1 $pn$ pairing correlations.  Bertsch and Zuo demonstrated\cite{bertsch10} that a spin triplet pairing condensate may occur in large nuclei, where the spin-orbit term tends to become smaller. Their study is based on a Woods-Saxon description for the mean field of the nucleons and a phenomenological contact interaction to generate the pairing correlations.

The present study tries to investigate the development of the spin-orbit term and $T=1$ as well as $T=0$ pairing-correlations in a model, 
which is based on a realistic nucleon-nucleon ($NN$) interaction and allows for a continuous change of the size of the nuclear system ranging from infinite nuclear matter to light nuclei. 

This model as well as the technique to determine the  pairing for spin-singlet, $T=1$, and  spin triplet, $T=0$, pairing are presented in section 2 of this paper. Results for isospin-symmetric infinite matter, the dependence of spin-orbit term, as well as the pairing gaps on the size of the system, and results based on self-consistent Hartree-Fock calculations are discussed in section 3. The conclusions from the present study as well as possible extensions of this work are summarized in the final section 4.

\section{Pairing in Nuclear Matter and Nuclei}
\subsection{Basis States}
One of the central goals of the present project is to develop a computational scheme, which allows to describe a smooth transition from infinite nuclear matter to finite nuclei. For that purpose the momentum eigenstates for free particles in a spherical box are considered as a common set of basis states for the single-particle states in nuclear matter as well as finite nuclei. These basis wave functions can be separated  
  in a radial  and an angular part,
\begin{equation}
\Phi_{iljm}(\vec r) = \langle \vec r \vert iljm\rangle = R_{il}(r)
{\cal Y}_{ljm} (\vartheta,\varphi)\,.\label{eq:basis}
\end{equation}
where ${\cal Y}_{ljm}$ represent the spherical harmonics including
the spin degrees of freedom by coupling the orbital angular momentum $l$ with
the spin to a single-particle angular momentum $j$.

The radial part $R_{il}$ of a particle moving free in a spherical
cavity with a radius $R$, the plane wave (PW)
basis~\cite{artur95,Montani:2004,erik02}, is then given by the
spherical Bessel functions,
\begin{equation}
R_{il}(r) = N_{il} j_l(k_{il}r),
\label{eq:PW_Radial}
\end{equation}
for the discrete momenta $k_{il}$, which fulfill
\begin{equation}
R_{il}(R)= N_{il}~j_l(k_{il}R) = 0 \ => \ j_l(k_{il}R) = 0\, .\label{eq:PW_condition}
\end{equation}
The normalization constant in Eq.~(\ref{eq:PW_Radial}) is given by
\begin{equation}
N_{il}=\left\{\begin{array}{ll}
i\pi \frac{\sqrt{2}}{\sqrt{R^3}}&\mbox{for}\,l=0\,,\\
\frac{1}{j_{l-1}(k_{il}R)} \frac{\sqrt{2}}{\sqrt{R^3}} &
\mbox{for}\,l>0\,.\end{array}\right.
\end{equation}
It ensures that the basis functions are orthogonal and normalized within the box,
\begin{equation}
\int_0^R d^3r~ \Phi^*_{iljm}(\vec r)\Phi_{i'l'j'm'}(\vec r) =
\delta_{ii'}\delta_{ll'}\delta_{jj'}\delta_{mm'}\,.
\end{equation}
 
In the mean-field approximation for infinite nuclear matter these basis states will be considered as the single-particle states of the system, occupying all states with momenta $k_{il}$ below a Fermi momentum $k_F$. If we represent the set of quantum numbers $iljm$ by letters $a$ or $b$ this leads to single-particle energies
\begin{equation}
\varepsilon_a = \frac{(\hbar k_{a})^2}{2M} + \sum_{b<F} \langle \Phi_a\Phi_b \vert V\vert \Phi_a\Phi_b \rangle\,,\label{eq:epsnucmat}
\end{equation}
where the sum is restricted to states $b$ with $k_{il}$ below the Fermi momentum $k_F$ under consideration, $M$ stands for the mass of the nucleon and $ \langle ab \vert V\vert ab \rangle$ are matrix elements of a realistic $NN$ interaction, which we will discuss below.

It is obvious that this treatment must be considered as an approximation to infinite nuclear matter: The single-particle energies $\varepsilon_a$ are not continuous but discrete with spacings depending on the size of the box $R$ and the density profile is quite different from homogeneous with $\rho (R) = 0$ due to the chosen boundary condition. In the next section we will show to which extent this approach represents features of homogeneous infinite matter.

The set of basis states $\Phi_{iljm}$ can also be used to diagonalize the Hamiltonian of a simple Woods Saxon potential, which consists out of the kinetic energy and a Woods Saxon potential of the form
\begin{equation}
U_{WS}(r) = \frac{U_0}{1 + e^{\left(\frac{r-r_0}{a_0}\right)}}\,.\label{def:uws}
\end{equation}
The parameter $a_0$ defining the skin-thickness has been fixed to $a_0 = 0.5$ fm$^{-1}$. A set of mean radii $r_0$ has been considered ranging from $r_0 = 2.5$ fm to $r_0 =20$ fm. In order to simulate quasi-nuclear systems in the range of $^{16}O$ for each value of $r_0$ a corresponding depth of the potential $U_0$ has been evaluated which yields a single-particle energy of zero for the first excited $s_{1/2}$ state. In a second set of Woods Saxon potential
the depth parameter $U_0$ has been fitted to ensure that the first excited $p$ states occur at zero energy. This should lead to single-particle wave functions, which describe quasi-nuclear systems close to $^{40}$Ca with a variety of radii ranging from a realistic size to quasi-nuclei, which are very weakly bound with single-particle wave functions close to plane waves.

The output of the calculations using these sets of Woods Saxon potentials are the resulting wave functions, which are expressed in terms of expansion coefficients, $ c^{WS}_{nil}$, 
\begin{equation}
\vert \Psi^{WS}_{nljm} \rangle = \sum_i c^{WS}_{nil} \vert  \Phi_{iljm} \rangle \,.\label{eq:expanws}
\end{equation}
Notice, that these expansion coefficients neither depend on the projection quantum number $m$ nor on the angular momentum $j$ since the Woods Saxon potential of Eq.(\ref{def:uws}) is purely central without any spin-orbit term. 

Assuming that these Woods Saxon wave functions can be considered to represent the Hartree-Fock single-particle wave functions of the quasi-nuclear system of a size determined by the radius parameter $r_0$ in Eq.(\ref{def:uws}) one may then evaluate the corresponding single-particle energies
\begin{eqnarray}
\varepsilon^{WS}_{nlj} & = & \langle \Psi^{WS}_{nljm}\vert \hat t \vert \Psi^{WS}_{nljm}\rangle + \label{eq:epsws}\\
&& \nonumber \sum_{(n'l'j'm')<F}
 \langle \Psi^{WS}_{nljm} \Psi^{WS}_{n'l'j'm'}\vert V \vert \Psi^{WS}_{nljm} \Psi^{WS}_{n'l'j'm'}\rangle \,.
\end{eqnarray}
In this equation, $\hat t$ stands for the kinetic energy and the sum is restricted to states below the Fermi surface $F$ for the quasi-nuclear system considered.
One should be aware that these single-particle energies will depend on the total angular momentum $j$ of the single-particle state considered, reflecting the effect of a spin-orbit term derived from the realistic NN interaction $V$. These values will be used below to explore the evolution of the spin-orbit term from the finite size of the quasi-nuclear system considered. 

The same set of basis states (\ref{eq:basis}) can also be used to represent the corresponding single-particle states resulting from self-consistent Hartree-Fock calculations
\begin{equation}
\vert \Psi^{HF}_{nljm\tau} \rangle = \sum_i c^{HF}_{nilj\tau} \vert  \Phi_{iljm\tau} \rangle \,.\label{eq:expanHF}
\end{equation}
Since we will restrict our studies in this work to a spherical description of nuclei, the expansion coefficients do not depend on the projection quantum number $m$. However, they will depend on the angular momentum $j$ and on the isospin $\tau$, if we account for the Coulomb interaction in $N=Z$ nuclei.

This basis of single-particles constrained to a spherical box can be compared to the basis of appropriate harmonic oscillator states, 
which is more popular in nuclear structure studies. The major advantage of the ``box basis'' as compared to the harmonic 
oscillator for the present study is the possibility to describe the transition from infinite matter to finite nuclei within 
the same set of basis states. Another advantage of the box basis is that it allows for a more realistic description of 
the continuum of single-particle states above the Fermi surface of finite nuclei. A major disadvantage of the box basis 
is the fact that typically a larger number of basis states is required to obtain a reliable representation in the box basis 
as compared to the oscillator basis. Another, more technical, disadvantage is the fact that for the evaluation of matrix elements 
of a realistic NN interaction a transformation from relative coordinates to a laboratory frame is required. The evaluation 
of this transformation in terms of vector brackets\cite{clement,bonats1,bonats3} required for the box basis is more demanding than the corresponding Talmi-Moshinsky transformation\cite{moshinsky:1959,talmi:1952} to be used in the case of oscillator states.

\subsection{Realistic NN Interaction}
It has already been mentioned that the box basis described above generally requires a large number of basis states in order to provide a reliable description of the nuclear wave functions. This is true in particular if one employs traditional realistic NN interactions like e.g. the Argonne V18\cite{v18} or the CDBonn potential\cite{cdbonn}, which contain strong short-range, respectively high-momentum components to reproduce the NN scattering data and the  deuteron observables with high accuracy. 

A possible way out of this problem is to consider an interaction model, which
separates the low-momentum (below a cut-off $\Lambda$) and high-momentum 
components of a realistic NN interaction by means of renormalization
techniques\cite{bogner01,bogner:2005,bogner:2007,bozek:2006}. If the cutoff
$\Lambda$ is chosen around $\Lambda$ = 2 fm$^{-1}$ the resulting low-momentum
interaction $V_{lowk}$ still describes the NN scattering data up to the pion
threshold and turns out to be independent of the underlying realistic 
interaction $V$. Since the high-momentum components, which correspond to the
short-distance behavior, of $V$ have been removed, the resulting $V_{lowk}$
does not produce significant short-range correlations, and  can be 
treated within the Hartree-Fock approximation\cite{bozek:2006}. 

In the present study we will use a $V_{lowk}$ interaction, which has been derived from the CDBonn interaction by means of the unitary-model-operator approach (UMOA)
\cite{suzuki:1982} using a cutoff $\Lambda$ = 2 fm$^{-1}$. Employing such a $V_{lowk}$ in nuclear structure calculations one does
not reproduce the empirical saturation features. The energy per nucleon in nuclear matter increases with density in a monotonic 
way\cite{bozek:2006,kuckei}. Calculations of finite nuclei predict radii which are much smaller than the empirical data\cite{review17}.
Similar features are observed if one just uses the two-body part of modern chiral interaction models\cite{chiral}. The way out of this problem is to include 3-nucleon interactions or a density-dependent two-body interaction which yields the empirical saturation point. Following the approach of van Dalen et al.\cite{erik02} the $V_{lowk}$ is supplemented in the Hartree-Fock calculations discussed below by a simple
contact interaction defined in the notation of the Skyrme
interaction
\begin{equation}
\Delta\mathcal{V} = \Delta\mathcal{V}_0 + \Delta\mathcal{V}_3,
\label{eq:contact}
\end{equation}
with
\begin{equation}
\Delta\mathcal{V}_0 = \frac{1}{4}t_0\left[2\rho^2-(\rho_n^2+\rho_p^2)\right]
\end{equation}
and
\begin{equation}
\Delta\mathcal{V}_3 = \frac{1}{24}t_3 \rho^{0.5}\left[ 2\rho^2-(\rho_n^2+\rho_p^2)\right],
\end{equation}
where $\rho_p$ and $\rho_n$  refer to the local densities for protons and neutrons
while the matter density is denoted as $\rho=\rho_p+\rho_n$. 
The parameters of the contact interaction are $t_0$ and 
$t_3$, which have been fitted
in such a way that HF calculations using $V_{lowk}$ plus the
contact term of Eq.~(\ref{eq:contact}) yield the empirical saturation point for
symmetric nuclear matter\cite{erik02}.

\subsection{Pairing Gap}
The analysis of pairing correlations of the present project is based on the method of evaluating self-consistent Greens functions
(SCGF)\cite{Dickhoff}. Within this framework one tries to determine the Greens function for two interacting nucleon in a nuclear medium or the corresponding $T$ matrix. Such calculations have been plagued  by so-called pairing instabilities, which are related to the
occurrence of quasi-bound two-nucleon states in the nuclear medium\cite{ramos1,frick05}. 

Recently, Rubtsova et al.\cite{rubts17,NM1} developed a formalism in which the two-particle Greens function is evaluated in terms of
discrete eigenvalues and eigenfunctions of a two-particle Hamiltonian. It includes particle-particle states ($pp$) but also the hole-hole states ($hh$) and corresponds to the Hamiltonian of the $pphh$ RPA
\begin{equation}
\left(\begin{array}{cc} H^0_{p} + V_{pp} & V_{ph} \\ -V_{hp} & H^0_h-V_{hh}\\ \end{array}\right) \,. \label{eq:pphhrpa}
\end{equation}
In this matrix $H^0_p$ represents the single-particle part of the $pp$ Hamiltonian, which means that it can be written
$$
H^0_p = \sum_{p1,p2} \left(\tilde\varepsilon_{p1} + \tilde\varepsilon_{p2}\right) \vert p_1p_2 \rangle\langle p_1p_2\vert\,,
$$
with the single-particle energies
$$
\tilde\varepsilon_p = \varepsilon_p - \varepsilon_F
$$
rescaled by the Fermi energy $\varepsilon_F$. The corresponding definition applies to the $hh$ single-particle Hamiltonian $H^0_h$ while the two-body part of the RPA Hamiltonian is defined in terms of matrix elements of the form e.g.
$$
V_{ph} \Longleftrightarrow \langle p_1p_2\vert V \vert h_1h_2\rangle\,,
$$
and similar for $V_{pp}$, $ V_{hp}$ and $V_{hh}$. If, for the moment, we restrict the discussion to $pp$ and $hh$ states of nuclear matter with center of mass momentum $K_{cm}=0$, the $pp$ states are identified as a two-particle state with momenta $\vec k$ and $-\vec k$ with $\vert \vec k\vert > k_F$. This means that the after a partial wave decomposition such a state is characterized by one wavenumber $k$ which is larger than the Fermi-momentum $k_F$ for the $pp$ states and smaller than $k_F$ for the hole-hole states. 

Rubtsova {\it et al.} demonstrated that after discretizing the momentum variable $k$ a nontrivial solution of the BCS equation for nuclear matter in a specific partial wave is signaled by a pair of complex conjugated eigenvalues for the corresponding $pphh$ RPA Hamiltonian. Since the Hamiltonian in Eq.~(\ref{eq:pphhrpa}) is non-hermitian but real, it may have pairs of complex  eigenvalues $E_\beta$ and $E_\beta^*$ with conjugated eigenfunctions $\vert \Phi_\beta \rangle$ and $\vert \Phi_\beta^*\rangle$.  In fact the imaginary part of these eigenvalues 
\begin{equation}
\vert \Im E_\beta \vert = \Delta (k_F)\,,\label{eq:imagval}
\end{equation}
with $\Delta (k_F)$  the pairing gap at the Fermi surface. Furthermore it has been observed that the wavefunction of the bound state is proportional to the function of the pairing gap
$$
\left| \langle k \vert \Phi_\beta \rangle\right| \sim \Delta(k)\,,
$$
which means that the bound-state of the $pphh$ RPA is rather close to the solution of the BCS gap equation, which can be written in the form
\begin{equation}
\left(\begin{array}{cc} \tilde H^0_{p} + V_{pp} & V_{ph} \\ V_{hp} & \tilde H^0_h + V_{hh}\\ \end{array}\right)\left(\begin{array}{c} \langle p \vert \chi \rangle \\ \langle h \vert\chi \rangle \\ \end{array}\right) = 0 \,. \label{eq:pphhbcs1}
\end{equation}
with
\begin{equation}
\tilde H^0_p = \sum_{k>k_F} 2\sqrt{\left(\varepsilon_k -\varepsilon_F\right)^2 + \Delta(k)^2}\vert k\rangle\langle k\vert \,.\label{eq:pphhbcs2}
\end{equation}
and a corresponding definition for the $hh$ part $\tilde H^0_h$. The solution of the homogeneous Eq.~(\ref{eq:pphhbcs1}) , $\chi$, is to be interpreted as
\begin{equation}
\left| \langle k \vert \chi \rangle\right| = \frac{\Delta(k)}{2\sqrt{\left(\varepsilon_k -\varepsilon_F\right)^2 + \Delta(k)^2}}\,.\label{eq:pphhbcs3}
\end{equation} 
In fact, the representation of the BCS equation for nuclear matter in Eqs.~(\ref{eq:pphhbcs1} - \ref{eq:pphhbcs3}) leads to a very efficient way for the solution of the non-linear BCS equation: Assume that $\Delta = 0$ and determine the eigenvalues of the matrix in Eq.~(\ref{eq:pphhbcs1}). A nontrivial solution for the gap function is only obtained, if the lowest eigenvalue is below zero energy (compare discussion of Eq.~(\ref{eq:imagval})). The complete function $\Delta(k)$ can be obtained from an iterative solution of Eqs.~(\ref{eq:pphhbcs1} - \ref{eq:pphhbcs3}) until the lowest eigenvalue occurs at zero energy. This formulation of the gap equation in terms of a bound state, and the gap function needed to shift this bound state to zero energy is close to the discussion of pairing instabilities in \cite{Dickhoff}.

This direct interpretation of the gap function $\Delta(k)$ defining the quasiparticle energies $E_k$ in terms of a bound state of the $pphh$ RPA Hamiltonian or the Hamiltonian in Eq.~(\ref{eq:pphhbcs1}) is only possible for the $pp$ and $hh$ states in nuclear matter with vanishing center of mass momentum. In the general case of finite nuclei, as well as in the basis $pp$ and $hh$ states we have to deal with two-particle states in the laboratory frame using single-particle quantum numbers as in Eq.~(\ref{eq:expanws}). Such two-particle states are coupled to total angular momentum $J$ and isospin $T$ leading to
\begin{equation}
\vert \Psi_{nljm} \Psi_{n'l'j'm'} \rangle_{JT}\,.\label{eq:twocoup}
\end{equation}
This means we do not separate pairing in different partial waves. For proton-proton 
or neutron-neutron pairing we will consider the case with $J=0$ and $T=1$, while for 
proton-neutron pairing in the isospin $T=0$ case, we will consider the case of $J=1$ 
and $T=0$. The occurrence of pairing can be identified either by an imaginary part of 
eigenvalues solving the $pphh$ RPA equations (\ref{eq:pphhrpa}). Assuming a constant pairing 
gap $\Delta$ for all single-particle states, one can adjust this pairing energy in the 
definition of the quasi-particle part of the Hamiltonian (\ref{eq:pphhbcs2}) until the 
lowest eigenvalue of Eq.~(\ref{eq:pphhbcs1}) occurs at zero energy.  Therefore the pairing gap defined in this way
is the minimal gap parameter to be used in the quasi-particle energies, defined in Eq.~(\ref{eq:pphhbcs2}), 
which is needed to stabilize the equation for the $pphh$ T-matrix against pairing instabilities. In fact, the results 
for the pairing deduced from Eqs.~(\ref{eq:imagval}) and (\ref{eq:pphhbcs1}) turn out to be rather similar.

It should be noted that in the case of the vacuum, i.e. there are no hole states and the single-particle energies are just kinetic energies, the diagonalization of the two-particle Hamiltonian in the box basis of Eq.~(\ref{eq:twocoup}) with $J=1$ and $T=0$ yields the energy and wavefunction of the deuteron with very good accuracy.

\section{Results and Discussion}
\subsection{Nuclear Matter}

\begin{figure}[!ht]
\begin{center}
\epsfig{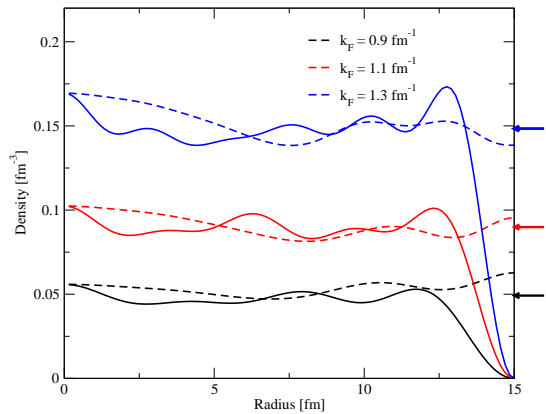} \caption{(Color online) The density profile for 
the model of nuclear matter calculated in a spherical box of radius $R$ = 15 $fm$, assuming various 
Fermi momenta $k_F$. The solid lines represent the density profiles calculated for plane waves with
boundary conditions according to Eq.~(\ref{eq:PW_condition}) while the dashed lines have been obtained using the alternating
boundary conditions discussed in the text. The arrows at the right axis indicate the densities 
for homogeneous matter according to Eq.~(\ref{eq:denskf})}
 \label{boxdens}.
\end{center}
\end{figure}

This section has been written to demonstrate the successes and limitations of the description of symmetric nuclear matter confined to a spherical box as described above. One of the obvious limitations is displayed in Fig.\ref{boxdens}, which shows the density profile of nuclear matter assuming various Fermi momenta. The solid lines are obtained if one defines the orthogonal basis for the plane waves according to Eq.~(\ref{eq:PW_condition}). The densities are not really constant but fluctuate around the densities
\begin{equation}
\rho = \frac{2}{3\pi^2}k_F^3\,, \label{eq:denskf}
\end{equation}
which one would expect for homogenous nuclear matter with the specific Fermi momentum $k_F$. In particular, the drop towards $\rho=0$ at  the boundary of the box, which is a direct consequence of Eq.~(\ref{eq:PW_condition}) seems to spoil the picture. Therefore it has been argued that one should employ a set of basis functions, with alternating boundary conditions: while the boundary condition of  Eq.~(\ref{eq:PW_condition}) is used for states with even orbital angular momentum $l$, the momenta for the states with odd $l$ are determined from the boundary condition that the derivative of the corresponding Bessel function $j_l(kr)$ vanishes at the boundary. 

Indeed the resulting density profiles, visualized in terms of the dashed lines in Fig.\ref{boxdens}, are a bit smoother and get rid of the drop at the maximal radius. Since, however, our main interest is devoted to the study of finite nuclei of variable size, we will stick to basis of  Eq.~(\ref{eq:PW_condition}) also for our studies of infinite matter as it leads to a sequence of orbital angular momenta, which is typical for finite systems.

\begin{figure}[!ht]
\begin{center}
\epsfig{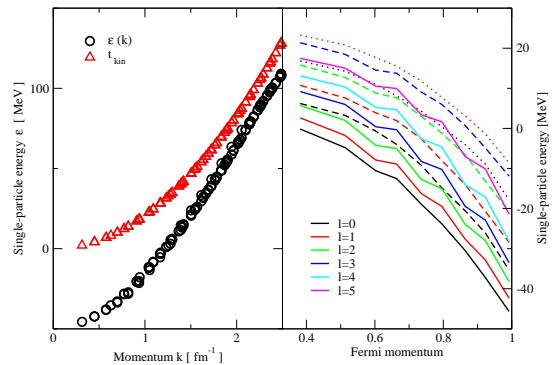} \caption{(Color online) Single-particle energies of
nuclear matter calculated in a spherical box of radius $R$ = 10 $fm$. The left panel shows the 
kinetic energies $t_{kin}$ and the single-particle energies $\varepsilon (k)$ calculated for a Fermi 
momentum $k_F$ =  1 $fm^{-1}$. The right panel shows values of $\varepsilon_{il}$ as a 
function of the Fermi momentum $k_F$. 
Results for various orbital angular momenta $l$ are identified by the color of the line. Solid, dashed
and dotted lines represent the values for the lowest 3 momenta according to the boundary conditions
	of Eq.(\ref{eq:PW_condition}).
 \label{nucepsi}}
\end{center}
\end{figure}

This can be seen from right panel of Fig.\ref{nucepsi}, which shows the single-particle energies for the lowest orbits calculated according to Eq.~(\ref{eq:epsnucmat}) as a function of the Fermi momentum of the system. One finds that these energies change rather smoothly with $k_F$. The lowest states are for orbital angular momenta $l=$ 0, 1 and 2, with the smallest momenta compatible with the boundary condition of  Eq.~(\ref{eq:PW_condition}). The next state, 
shown in terms of a dashed line, has again orbital angular momentum $l=0$ and can be regarded as  a first excited $s$-state in a finite system. It should be noticed that there is no spin-orbit splitting in the single-particle spectrum, if the radius of the box $R$ is chosen to be sufficiently large.
This applies to our choice of $R=10$ fm, which we will consider here and in the following discussion. 

The left panel of Fig. \ref{nucepsi}  shows the single-particle spectrum  as a function of the momentum for nuclear matter with $k_F$=1 fm$^{-1}$ together with the kinetic energy spectrum. The dependence on $k$ is rather smooth and could be parameterized in terms of a parabola defining an effective mass as it is often done in the literature. One must be aware, however, that the momenta compatible with the boundary conditions of Eq.~(\ref{eq:PW_condition}) are not equally spaced, as one can also see from the kinetic energies of these states. 

\begin{figure}[!ht]
\begin{center}
\epsfig{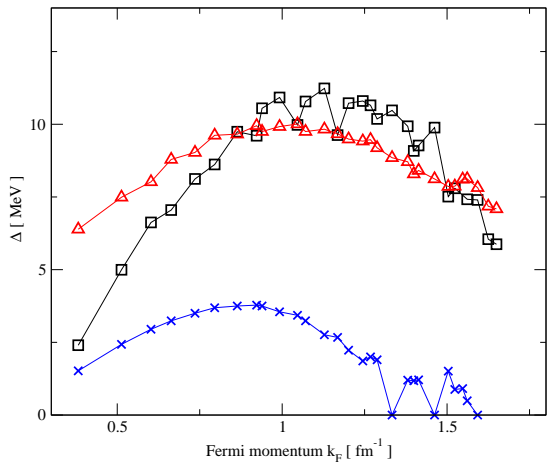} \caption{(Color online) Pairing gaps $\Delta$ 
calculated for nuclear matter with various Fermi momenta $k_F$. The triangles represent results 
	for $T=0$ pairing obtained from Eq.~(\ref{eq:pphhbcs1}) while the square boxes represent the corresponding 
	imaginary parts of the $pphh$ RPA eigenvalues (see Eq.~(\ref{eq:imagval})). Results for isospin $T=1$,
	derived from Eq.~(\ref{eq:pphhbcs1}) are 
shown in terms of x-signs. Note that the single-particle energies have been approximated by kinetic
energies.
 \label{deltapn12}}
\end{center}
\end{figure}

Results for pairing gap in nuclear matter are presented in Fig.\ref{deltapn12}. For the case of proton-neutron pairing the energy gaps resulting from the imaginary part of complex eigenvalues for the $pphh$ RPA (see Eq.~(\ref{eq:imagval})) are visualized by black squares while those resulting from Eq.~(\ref{eq:pphhbcs1}) are represented by triangles. The agreement between the predictions from these two approaches is not as close as in the case of the corresponding equations discussed by Rubtsova et al.\cite{rubts17}. This is not really astonishing as the studies of Ref. \cite{rubts17} were restricted to pairing in two-particle states with zero center of mass momentum and a well defined partial wave for the relative motion, while in the present investigation the center of mass momentum as well as the partial wave of relative motion are not well defined quantum numbers. As a consequence one typically obtains more than one pair of RPA eigenstates with complex eigenvalues. The magnitude predicted for the the proton-neutron pairing gap in nuclear matter as well as the dependence of the gap on the Fermi momentum are in reasonable agreement with each other as well as with the corresponding results obtained e.g. in \cite{rubts17}. Therefore in the following we will consider Eq.~(\ref{eq:pphhbcs1}) to define the pairing gap.

Fig.\ref{deltapn12} also displays results for the pairing gap $\Delta$ in the case of neutron-neutron or proton-proton pairing. These results for the $T=1$
pairing gap are considerably smaller than those for $T=0$ pairing. This in agreement of earlier studies in nuclear matter and supports the expectation that the more attractive NN interaction in $T=0$ channels should produce stronger correlation effects than in the $T=1$ case. The fluctuations for the pairing gap in the $T=1$ channel at larger densities reflects the fact that the existence or non-existence of a pairing solution is quite sensitive to the energy-spectrum of the single-particle energies in the case of a weak pairing effects.

\begin{figure}[!ht]
\begin{center}
\epsfig{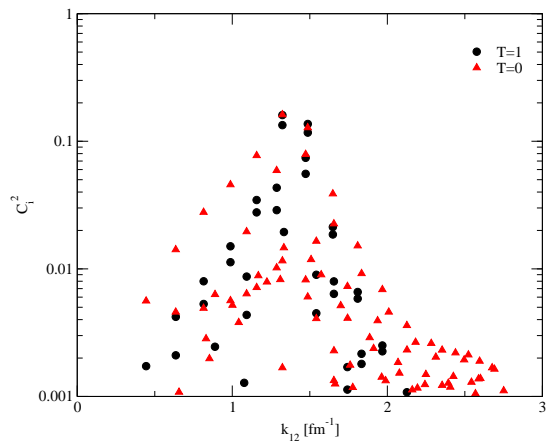} \caption{(Color online) Expansion coefficients of
the bound states for $T=0$ and $T=1$ pairing in nuclear matter with $k_F$ = 1 fm$^{-1}$. 
The states of the two-particle basis are represented by the average momentum $k_{12}$ 
defined in Eq.~(\ref{eq:k12}).
 \label{wavesnucmat}}
\end{center}
\end{figure}

In comparing the pairing in $T=0$ and $T=1$ one should be aware that the quasi-bound states in these two channels show rather different features. This can be seen from Fig.\ref{wavesnucmat}, which displays the expansion coefficients of the lowest eigenstates of the Hamiltonian in Eq.~(\ref{eq:pphhbcs1}). In this figure the expansion coefficients $c_i$ of the eigenstates are identified by a typical value for the momentum
\begin{equation}
k_{12} = \sqrt{k_1^2 + k_2^2}\,, \label{eq:k12}
\end{equation}
where $k_1$ and $k_2$ identify the momenta of the two  single-particle states in the basis of $pp$ and $hh$ states. For a Fermi-momentum $k_F$= 1 fm$^{-1}$ this implies that $pp$ and $hh$ states very close to the Fermi energy are identified by $k_{12} = \sqrt{2}$ fm$^{-1}$, which is the value around which 
the maximal values for $c_i^2$ occur in the $T=1$ (dots) as well as in the $T=0$ (triangles)  case. One also sees, however, that non-negligible values for $c_i$ occur at larger values of $k_{12}$ in the case of $proton-neutron$ states, which is not true for the case of nucleons with the same isospin. One should be aware that the logarithmic scale in this figure suppresses the value for most expansion coefficients, which are of the order of 1000 in each channel for the present study.

\begin{figure}[!ht]
\begin{center}
\epsfig{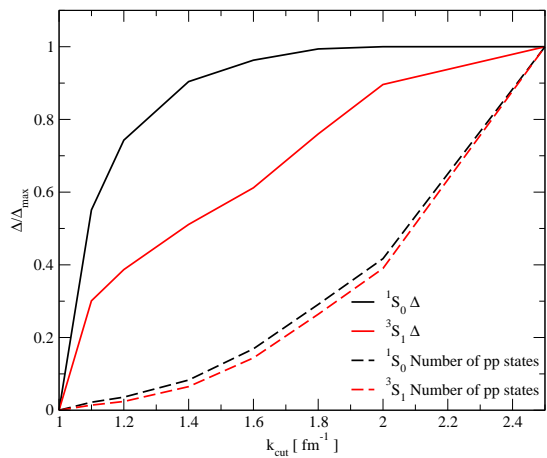} \caption{(Color online) Value for the pairing gap
$\Delta$ for $T=0$ and $T=1$ pairing as a function of the cut-off momentum for the particle states are
represented by solid lines while the number of basis states is indicated by the corresponding dashed
lines. All value are normalized with respect to the result, both gap and number of basis states, at the maximal cut-off $k_{cut}$ of 2.5
fm$^{-1}$.
 \label{deltaofkmax}}
\end{center}
\end{figure}

These expansion coefficients reflect the fact that the strong tensor components of a realistic NN interaction in the $^3S_1-^3D_1$ interaction yield high-momentum components in the quasi-bound states for $T=0$, which are absent in the case of $T=1$ states. This feature is also reflected by the analysis displayed in Fig.\ref{deltaofkmax}, which shows the dependence of the pairing gap $\Delta$ on the cut-off in the single-particle momenta of particle states, which are considered for the basis of particle-particle states. Note that in all the studies discussed in this project we consider a basis of single-particle states up to a maximal momentum of $k$ = 2 fm$^{-1}$. This turned out to be sufficient for the low-momentum representation of a realistic interaction $V_{lowk}$, which is used here. Larger cut-offs will be required if one considers traditional realistic NN interactions.

Inspecting Fig.\ref{deltaofkmax} one finds that a cut-off larger than 2 fm$^{-1}$ is needed in the $pp$ states to obtain stable results for $T=0$ while a lower cut-off is sufficient for the case of $T=1$ pairing.

\begin{figure}[!ht]
\begin{center}
\epsfig{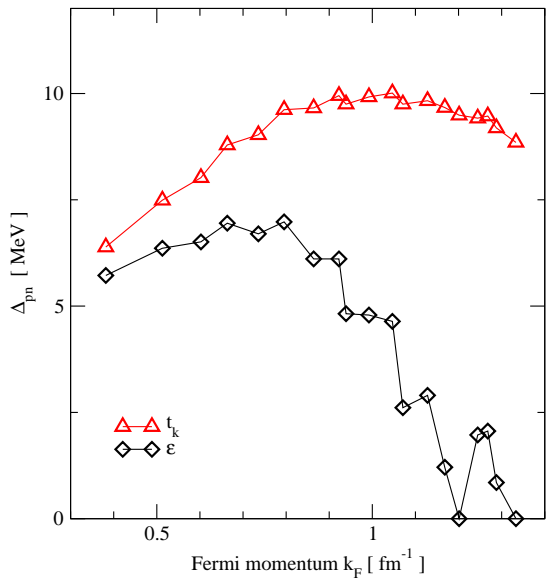} \caption{(Color online) Pairing gaps $\Delta$ 
calculated for nuclear matter with various Fermi momenta $k_F$. The symbols represent results 
	for $T=0$ pairing obtained from Eq.~(\ref{eq:pphhbcs1}). while the diamonds represent results obtained
for the corresponding HF single-particle energies, the triangles were obtained approximating these
energies by kinetic energies $t_k$. 
 \label{deltapn3}}
\end{center}
\end{figure}

All the studies discussed so far in this subsection have used a spectrum of kinetic energies $\varepsilon_k$ for the single-particle
states in Eq.~(\ref{eq:pphhbcs2}). This has been done to allow for a direct comparison with results obtained in \cite{rubts17}. 
The effect of a more realistic single-particle spectrum calculated according to Eq.~(\ref{eq:epsnucmat}) is displayed in Fig.~\ref{deltapn3}. The quenching of the pairing gap $\Delta$ with the single-particle spectrum, which in nuclear matter calculation is often represented by an effective mass, $m^*<m$, has also been observed in earlier studies. 

\subsection{Effects of finite size}

\begin{figure}[!ht]
\begin{center}
	\epsfig{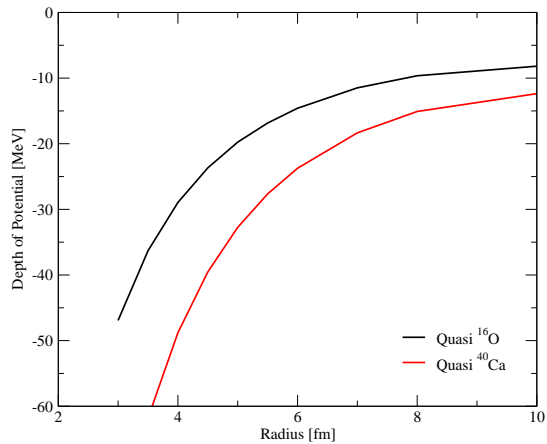} \caption{(Color online) Depth of Woods Saxon potentials $U_0$, defined in Eq.(\ref{def:uws}) to simulate quasi-nuclear systems corresponding to $^{16}O$ and $^{40}Ca$ for various radii as discussed in the text.
 \label{woodsdepth}}
\end{center}
\end{figure}

A central aim of this study is to explore the effects of the finite size of nuclei on the occurrence of pairing phenomena. 
For that purpose two sets of parameters of a simple Woods Saxon potential (Eq. \ref{def:uws}) have been generated. 
The first tries to simulate quasi-nuclei close to the double magic nucleus $^{16}O$. For a set of radial parameters $r_0$ in Eq.(\ref{def:uws}) the  depth
of the potential, $U_0$, has been adjusted such the energy of the first excited $s_{1/2}$ state occurs at zero energy. In a second set, which will be referred to as quasi $^{40}Ca$ nuclei, the depth has been adjusted 
to obtain the first excited $p$ state at zero energy. Note that there is no spin-orbit term in the potential of Eq.(\ref{def:uws}), which means that the
energies of $p_{3/2}$ and $p_{1/2}$ states are degenerated. The results for the depth of these sets of potentials are displayed in Fig.\ref{woodsdepth}.

\begin{figure}[!ht]
\begin{center}
\epsfig{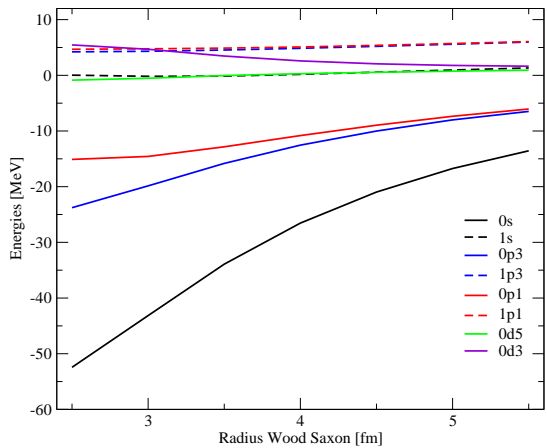} \caption{(Color online) Single-particle energies of low-lying states in quasi-nuclear systems corresponding to $^{16}O$ of different size.
 \label{quasioepsi}}
\end{center}
\end{figure}

Using the single-particle wavefunctions from such Woods Saxon potentials one can then calculate the corresponding mean-field energies following Eq.~(\ref{eq:epsws}). Results for the energies of the low-lying single-particle states using the Wood Saxon wavefunction of the quasi-$^{16}O$ nuclei are displayed in Fig.\ref{quasioepsi}.
 
From this figure one can clearly see that the single-particle energies for the states with bound-state wavefunctions are getting more attractive with decreasing size of the nucleus, while the energies for single-particle states with unbound Wood Saxon wavefunctions, as $0d_{5/2}$, $0d_{3/2}$ and the $1p$ states, are rather insensitive to the size of the nuclear system and stay around zero energy. This enhancement of the shell-structure for the bound states with decreasing size is, what one expects for quantum systems.

Special attention shall be paid to the sensitivity of the spin-orbit term in the single-particle energies on the size of the quasi-nuclei. The spin-orbit term in nuclei occurs in a  very natural way within the framework of a relativistic mean-field approach\cite{serot86,review17}. In such models the nuclear binding results from a balance between a scalar field and a repulsive vector field. Reducing the resulting Dirac equation to an equivalent Schr\"odinger equation, one obtains a strong spin-orbit term, which is proportional to the radial derivative of difference between the scalar and the vector field. Therefore, in non-relativistic mean-field studies the spin-orbit term is added also proportional to the radial derivative of the single-particle potential\cite{ring}.

The present study does not consider any phenomenological spin-orbit term. The Wood Saxon wavefunction are identical e.g. for the $0p_{3/2}$ and $0p_{1/2}$ states. Therefore the spin-orbit term showing up in the $0p$ and $0d$ states of the quasi-nuclear systems refering to $^{16}O$ in Fig.\ref{quasioepsi}
originates solely from the realistic NN interaction. 

\begin{figure}[!ht]
\begin{center}
\epsfig{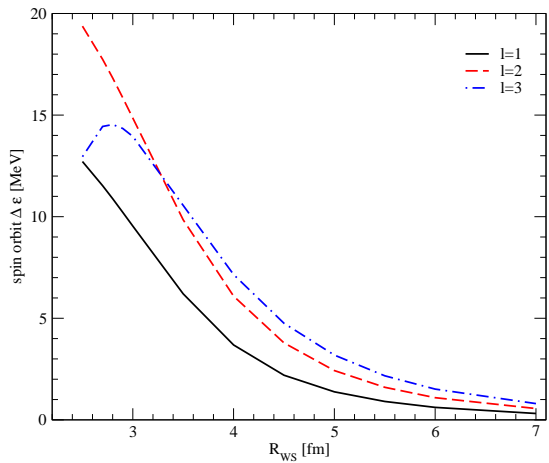} \caption{(Color online) Spin-orbit term defined by the difference in energy between single-particle states with angular momentum $j=l-1/2$ and $j=l+1/2$ for orbital angular momenta $l=1$, 2 and 3 for quasi-nuclei $^{40}Ca$ of various size.
 \label{lsterm}}
\end{center}
\end{figure}

This is also true for the spin-orbit energies 
$$
\varepsilon_{ls} = \varepsilon_{j=l-1/2} - \varepsilon_{j=l+1/2}\,,
$$
displayed in Fig. \ref{lsterm} for the quasi-nuclear system of various sizes, which corresponds to $^{40}Ca$. One can see that 
the spin-orbit term decreases with increasing size of the system and disappears in the limit of infinite nuclear matter. As already mentioned this spin-orbit term is due to the NN interaction. It turns out that the spin-orbit splitting is mainly due to interaction in partial waves with total spin $S=1$ and orbital angular momentum $L=1$ for the relative motion of the interacting nuclei. If the interaction in the $^3P_0$, $^3P_1$ and $^3P_2-^3F_2$ channels is ignored, spin-orbit energies of 0.61 MeV, 0.87 MeV and  0.70 MeV are obtained for $l=1$, 2 and 3 in the case of $^{40}Ca$ and a Wood Saxon radius of $r_0=3.5$ fm. This is much smaller than the corresponding energies of 6.20 MeV, 9.85 MeV, and 10.56 MeV, which one obtains for the interaction in all partial waves and displayed in Fig.\ref{lsterm}.

\begin{figure*}[!ht]
\begin{center}
\epsfig{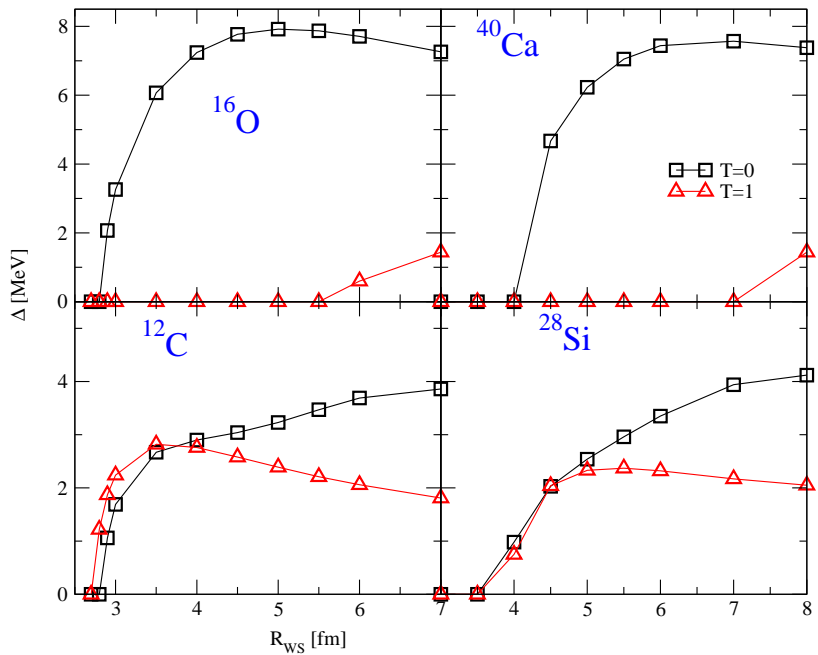} \caption{(Color online) Pairing gaps $\Delta$ calculated
for various quasi-nuclear systems are presented as a function of the radius of the Wood Saxon potential. 
While for the two panels on the left side ($^{16}O$ and $^{12}C$)  the wavefunctions and energies resulting
from the $^{16}O$ set have been used, the results for the other panels have been obtained using the $^{40}Ca$ sets.
Results for $T=0$  and $T=1$ pairing are displayed in terms of squares and triangles, respectively.
 \label{woodsdelta}}
\end{center}
\end{figure*}

Results for the pairing gap of quasi-nuclei derived from a solution of Eq.~(\ref{eq:pphhbcs1}) are shown in Fig.~\ref{woodsdelta} as a function of the radius 
of   the underlying Wood Saxon potential. It should be noted that the results for $^{16}O$ as well as for the open shell nucleus $^{12}C$ have been calculated using the wavefunctions and single-particle energies for $^{16}O$ discussed above (see e.g. Fig.~\ref{quasioepsi}). In the case of $^{12}C$ One would prefer of course to use single-particle energies evaluated directly for this nucleus. If, however, one calculates those single-particle energies according to Eq.~(\ref{eq:epsws}), assuming the states of the $0s_{1/2}$ and $0p_{3/2}$ shells to be occupied, one obtains a level inversion in the sense that the single-particle energy of the $0p_{1/2}$ occurs below the $0p_{3/2}$ energy. This is a consequence of the assumption of spherical approximation for the open shell-nucleus $^{12}C$. In a more realistic calculation one would allow for a deformation of the nuclei. This is beyond the scope of the present investigation and should be considered in a next step.

The situation is essentially identical for the open shell nucleus $^{28}Si$ using the Wood Saxon basis as well as in the Hartree-Fock calculations discussed in the next subsection. Therefore the pairing gaps displayed in Fig.\ref{woodsdelta} for $^{28}Si$ have been calculated using single-particle states as determined for $^{40}Ca$.

The dependence of the calculated pairing gaps on the size of the system turns out to be very similar for the closed shell nuclei $^{16}O$ and $^{40}Ca$ presented in the upper half of the Fig.\ref{woodsdelta}. In the case of $T=1$ pairing, one can observe that the pairing gap evaluated for homogeneous matter decreases and disappears when h the radius of the nuclear system decreases. The energy spacing between particle- and hole-states is getting large at small and realistic sizes of these nuclei and prevents the occurrence of a pairing state in this channel.

The situation is different for $T=0$ pairing in the closed shell nuclei. The proton-neutron interaction is more attractive and collects contributions to the collective states over a larger range (see discussion of Figs.\ref{wavesnucmat} and \ref{deltaofkmax} above). This may explain that the pairing gap in the $T=0$ case remains large over a wide range of nuclear sizes and drops to zero only at rather small values for the radius. The small enhancement of the $T=0$ pairing gap at medium values of $R_{WS}$ could be caused by larger matrix elements of the interaction for states which are more localized.

The situation differs for the open-shell nuclei $^{12}C$ and $^{28}Si$ displayed in the lower part of Fig.\ref{woodsdelta}. 
Since the energy gap between particle- and hole-states is smaller for these open shell nuclei than for the closed-shell systems discussed before, the pairing gap determined for $T=1$ remains down to rather low radii of the nuclear system. On the other hand, the pairing gap calculated for $T=0$ is smaller by a factor of 2 as compared to typical values for the closed shell nuclei. This reduction is mainly due to the spin-orbit splitting. In the case of $^{16}O$ the two-particle configuration $\vert 0p_{3/2}, 0p_{1/2}, J=1, T=0\rangle$ provides an important contribution to the quasi-bound state of Eq.~(\ref{eq:pphhbcs1}), which defines the pairing gap. In the case of $^{16}O$ this configuration belongs neither to the particle-particle nor to the hole-hole states, which reduces the attraction of the quasi-bound states and consequently requires a small gap parameter for the stabilization. It is remarkable that the gaps required for $T=1$ and $T=0$ are very close to each other for the open-shell nuclei.
  
\subsection{ Hartree-Fock approach}

\begin{figure}[!ht]
\begin{center}
\epsfig{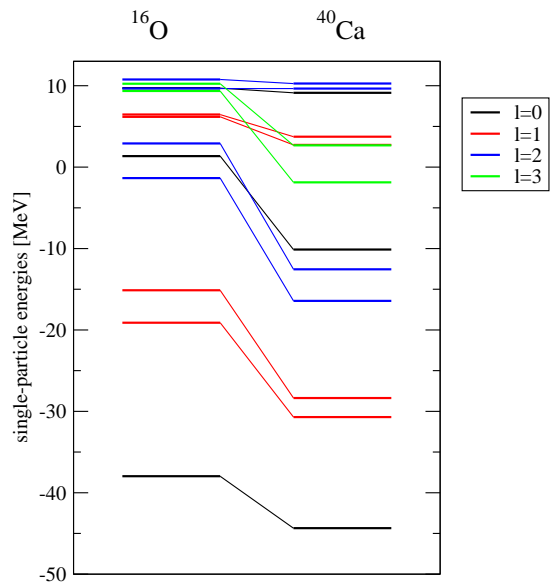} \caption{(Color online) Level scheme for low-lying single-particle energies for protons in $^{16}O$ and $^{40}Ca$ obtained in Hartree-Fock approximation. \label{hfeps}}
\end{center}
\end{figure}

After discussing the change of the pairing gap as a function of the size of the nuclear system, it is the aim of this subsection to provide estimates for $T=0$ and $T=1$ pairing for realistic examples of finite nuclei. For that purpose the realistic $V_{lowk}$  interaction has been supplemented by a density-dependent two-body interaction defined in Eq.~(\ref{eq:contact}) and the Coulomb interaction. The parameters of the contact interaction were adjusted in Hartree-Fock calculations to reproduce the binding energies and radii of the charge distribution of $^{16}O$ and $^{40}Ca$. 

Results of single-particle energies for the proton bound states and a few low-lying states in the continuum are displayed in Fig.\ref{hfeps}. These single-particle energies as well as the corresponding wave functions are employed to evaluate the pairing gaps according to Eq.~(\ref{eq:pphhbcs1}). As discussed above, the results of the Hartree-Fock calculations for $^{16}O$ and $^{40}Ca$ have also been used for the study of the open shell nuclei $^{12}C$ and $^{28}Si$. Results of such calculations are listed in Table \ref{table:pair}.

\begin{table}
\begin{center}
\begin{tabular}{|c|c c|}
\hline
& \ {T=1} \ & \ T=0 \ \\
\hline
$^{12}C$\ & 2.51 & 2.31 \\
$^{16}O$\ & 0.0 &  2.35 \\
$^{28}Si$\ & 1.78 & 1.10 \\
$^{40}Ca$\ & 0.0 & 3.30 \\
\hline
\end{tabular}
\end{center}
\caption{\label{table:pair}
Pairing gaps $\Delta$ for isospin $T=1$ and $T=0$ derived from Eq.~\protect{\ref{eq:pphhbcs1}} using Hartree-Fock wavefunctions and energies. All entries are in MeV.
}
\end{table}

These results demonstrate for the case of $T=1$ pairing that the energy gap in closed shell nuclei is too large to allow a formation of pairing correlations between like nucleons. For open-shell nuclei the proton-proton and neutron-neutron interaction are strong enough to support the formation of paired states.
The estimates for the pairing gap are in qualitative agreement with phenomenological studies. This supports the strategy of the present investigation, to relate
the pairing phenomenon to the stabilization of the $pphh$ RPA equation, defining the spectral distribution of the two-particle Greens function. 

In the case of 2-particle states with $T=0$ one obtains non-vanishing pairing gaps for closed-shell nuclei as well as open-shell nuclei. The two-nucleon interaction is stronger in this channel and collects contributions from states well below and high above the Fermi surface. Therefore the collective state of the $pphh$ RPA Hamiltonian is not as sensitive to details of the single-particle spectrum very close to the Fermi surface. This implies that one cannot expect a clear signal for the formation of $T=0$  pairing from the mass of neighbored nuclei comparable to the odd-even mass staggering in the case of $T=1$.
Instead, the $T=0$ two-nucleon correlations are responsible for a depletion of hole-state occupations, which is not restricted to the valence shell.

\section{Conclusions}
In order to investigate the dependence of pairing correlations on the size of the nuclear system a set of quasi-nuclear systems
with equal number of protons, $Z$, and neutrons, $N$, in the mass range from $A=12$ to $A=40$ has been considered. 
Constructing corresponding
quasi-nuclei with different sizes the change of pairing correlations in the transition from homogeneous infinite matter to finite
nuclei has been explored. 

In this study one observes the development of the spin-orbit term in the mean field of nuclei. In the present framework of 
non-relativistic nuclear structure calculations using realistic $NN$ interactions this spin-orbit term is mainly generated from
the interaction of two nucleons in partial waves with spin $S=1$ and an orbital angular momentum for the relative motion of 
$L=1$. The spin-orbit term disappears in the limit of homogeneous matter and 
provides results for the splitting of single-particle 
energies, $\varepsilon_{j=l-1/2}-\varepsilon_{j=l+1/2}$, for quasi-nuclei of realistic size, which are in qualitative agreement
with empirical data. 

It is this spin-orbit splitting, which suprresses the proton-neutron pairing in open shell nuclei drastically. While in infinite
nuclear matter the pairing gaps calculated for $T=0$ are typically larger by a factor of 4 compared to the $T=1$ pairing gap,
the gap calculated for open shell nuclei of finite size are slightly smaller for $T=0$ as compared to $T=1$. 

The situation is different for the study of the closed shell-nuclei $^{16}O$ and $^{40}Ca$: The large gap in the single-particle
energies around the Fermi-energy prevents the occurence of $T=1$ pairing, while a non-trivial solution of the gap equation is
obtained for $T=0$. This behaviour reflects another qualitative difference between $T=0$ and $T=1$ pairing. While the bound state
of $T=1$ states is dominated by particle-particle ($pp$) and hole-hole ($hh$) configurations of states rather close to the 
Fermi level, the residual interaction in the $T=0$ channel leads to important contributions from configurations, which are far
away from the Fermi level. This implies that one should not expect a sudden breakdown of $pn$ correlations 
comparing nuclei with
$Z=N$ to those with $Z=N\pm 1$. This may explain, that emprical data do not show a staggering of binding energies comparable to
the odd-even mass staggering indicating $T=1$ pairing. 

The $T=0$ pairing states derived from a realistic $NN$ interaction collect contributions from various major shells. Therefore one
should not expect a sell-model calculation, which is resticted to configurations within one major shell, 
to show the corresponding 
correlations.

The studies presented in this manuscript assume spherical symmetry of the nuclei. This should be sufficient
for the study of qualitative features in the transition from infinite matter to finite nuclei. For a more realistic study of
pairing correlations in finite nuclei, however, one should consider the effects of intrinsic deformation.  

\section{Acknowledgments}
This project has been supported by the DFG grant MU 705/10-2 and by MINECO (Spain) through Grant No. FIS2017-87534-P.

\end{document}